\documentclass[a4paper,11pt]{article}

\usepackage{jinstpub} 

\title{\boldmath Positrons vs electrons channeling in silicon crystal: energy levels, wave functions and quantum chaos manifestations}

\author[a,b]{N.F. Shul'ga,}
\author[c,1]{V.V. Syshchenko,\note{Corresponding author.}}
\author[c]{A.I. Tarnovsky,}
\author[c]{I.I. Solovyev}
\author[d]{and A.Yu. Isupov}

\affiliation[a]{Akhiezer Institute for Theoretical Physics of the NSC ``KIPT'',\\Akademicheskaya Street, 1, Kharkov 61108, Ukraine}
\affiliation[b]{V.N. Karazin National University,\\Svobody Square, 4, Kharkov 61022, Ukraine}
\affiliation[c]{Belgorod State University,\\Pobedy Street, 85, Belgorod 308015, Russian Federation}
\affiliation[d]{Laboratory of High Energy Physics (LHEP), Joint Institute for Nuclear Research (JINR),\\Dubna 141980, Russian Federation}

\emailAdd{syshch@yandex.ru}

\abstract{The motion of fast electrons through the crystal during
axial channeling could be regular and chaotic. The dynamical
chaos in quantum systems manifests itself in both statistical
properties of energy spectra and morphology of wave functions of
the individual stationary states. In this report, we investigate
the axial channeling of high and low energy electrons and
positrons near [100] direction of a silicon crystal. This case is
particularly interesting because of the fact that the chaotic
motion domain occupies only a small part of the phase space for
the channeling electrons whereas the motion of the channeling
positrons is substantially chaotic for the almost all initial
conditions. The energy levels of transverse motion, as well as the
wave functions of the stationary states, have been computed
numerically by the method presented at previous RREPS. Note that
the potential of the elementary cell in (100) plane of silicon
crystal possesses the symmetry of the square. The group theory
methods had been used for classification of the computed
eigenfunctions and identification of the non-degenerate and doubly
degenerate energy levels. The channeling radiation spectrum for
the low energy electrons has been also computed.}


\arxivnumber{1234.56789} 

\proceeding{XII$^{\text{th}}$ International Symposium <<Radiation from Relativistic Electrons in Periodic Structures>>\\
  September 18-22, 2017\\
  Hamburg, Germany}

\begin{document}
\maketitle
\flushbottom

\section{Introduction}
\label{sec:intro}

The fast charged particles incident onto the crystal under a small
angle to any crystallographic axis densely packed with atoms can
perform the finite motion in the transverse plane; such motion is
known as the axial channeling \cite{AhSh,AhSh2,Ugg}. The particle
motion in the axial channeling mode could be described with a good
accuracy as the one in continuous potential of the atomic string,
i.e. in the potential of atoms averaged along the string axis.
During motion in this potential the longitudinal particle momentum
$p_\parallel$ is conserved, so the motion description is reduced
to two-dimensional problem of motion in the transversal plane.
This transverse motion could be substantially quantum \cite{AhSh}.

From the viewpoint of the dynamical systems theory, the channeling
problem is interesting because the particle's motion could be both
regular and chaotic. The quantum chaos theory \cite{8,9,10,11,13,Reichl} predicts qualitative
difference for the particle's motion features in the cases of its regular and chaotic motion in the classical limit. These differences concerning both the wave functions of the individual stationary states and the statistics of the energy levels series have been demonstrated for the channeling electron in the semiclassical domain \cite{Poverh.2015, NIMB.2016}, where the energy levels density is high. The aim of the present report is to consider the opposite case when the total number of energy levels in the potential well is small. The energy levels and wave functions of the stationary quantum
states are computed in the present paper as well as the radiational transitions between them.

\section{Method and potential wells}
\label{sec:method}

The electron transversal motion in the atomic string continuous potential is described by the two-dimensional Schr\"odinger equation with Hamiltonian
\begin{equation}
    \label{Hamiltonian}
    \hat{H} = - \frac{\hbar^2}{2 E_\parallel / c^2} \left[ \frac{\partial^2}{\partial x^2} + \frac{\partial^2}{\partial y^2} \right] + U(x, y)
\end{equation}
and the value $E_{\parallel} / c^2$ (here $E_{\parallel} = (m^2
c^4 + p_{\parallel}^2 c^2)^{1/2}$) instead of the particle mass
\cite{AhSh}. The Hamiltonian eigenfunctions as well as the
transverse energy $E_\perp$ eigenvalues are found numerically using
the so-called spectral method \cite{3}. For the channeling problem
it has been applied for the first time in \cite{Dabagov3}. The
details of the method have been described previously in
\cite{5,6,RREPS15}.

\begin{figure}[htbp]
\centering
\includegraphics[width=0.49\textwidth]{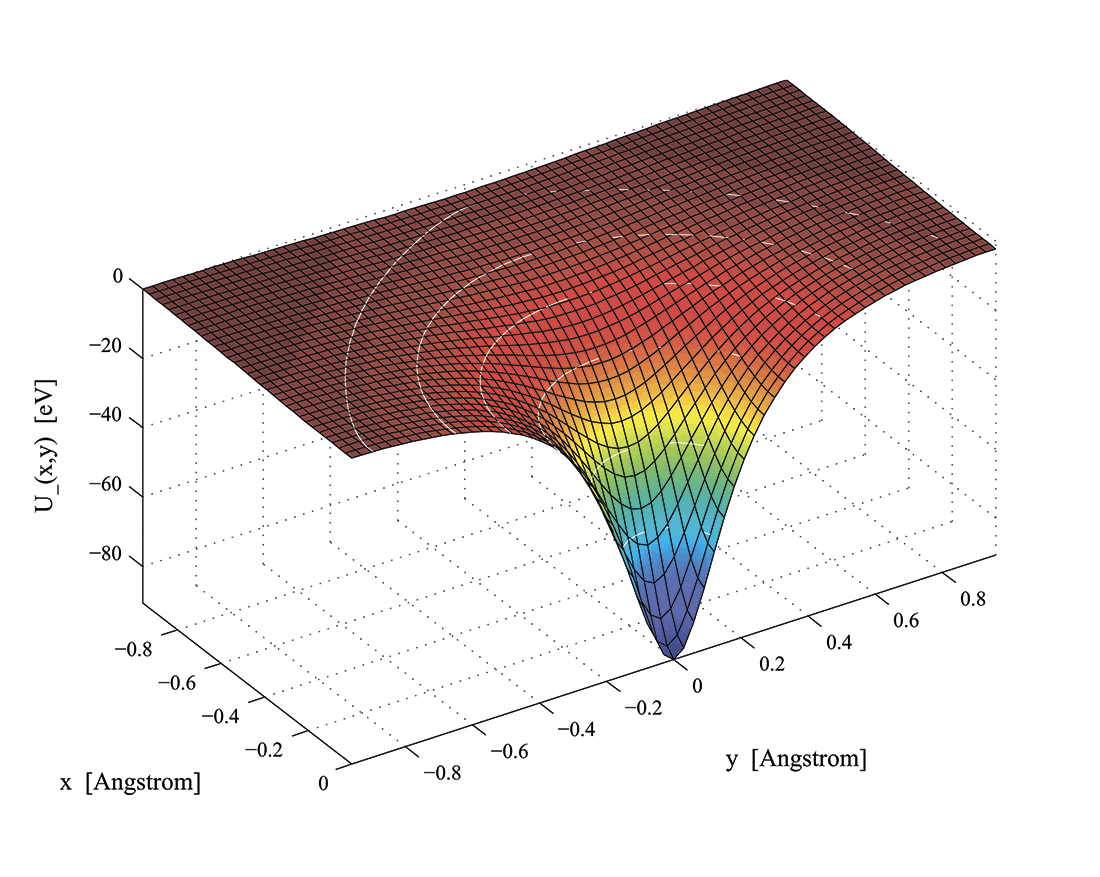} \
\includegraphics[width=0.49\textwidth]{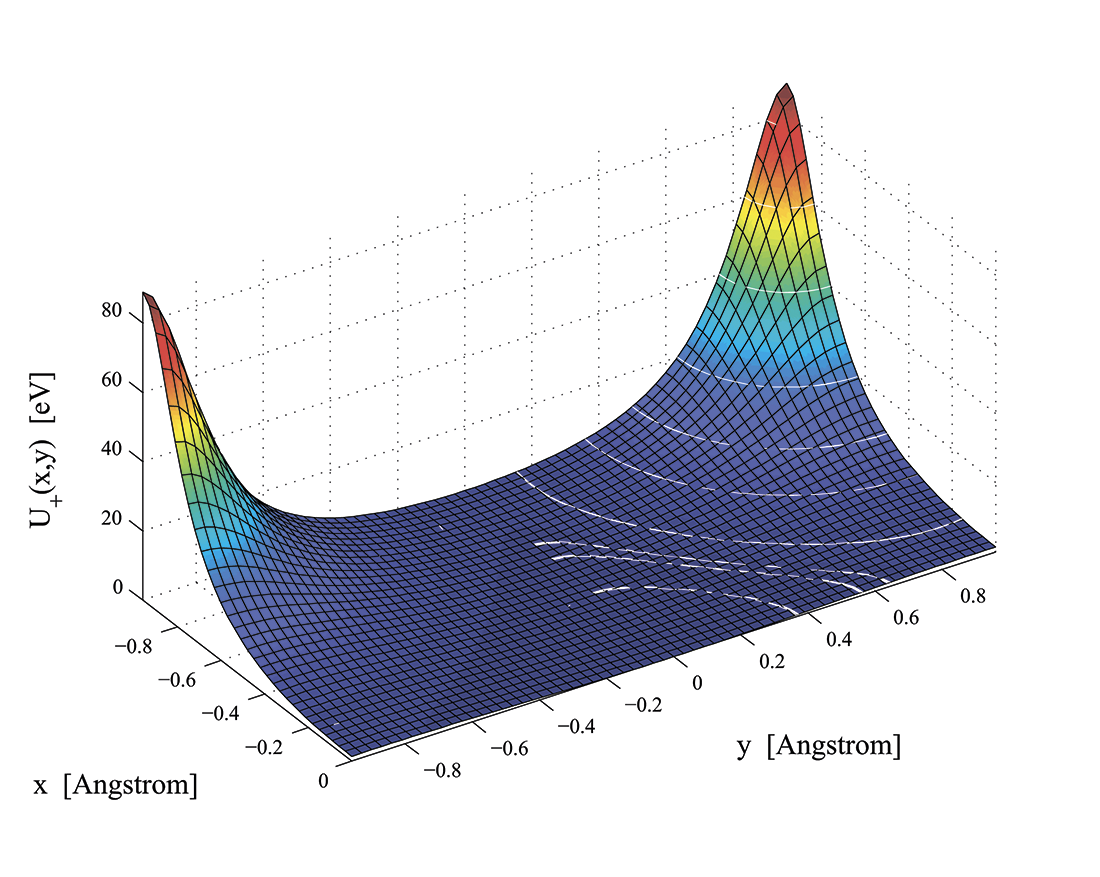}
\caption{\label{fig.potentials} The potentials (\ref{U.el}) and (\ref{U.pos}).}
\end{figure}

Here we consider the particle's motion near direction of the
atomic string $[100]$ of the Si crystal. The continuous potential
could be represented by the modified Lindhard potential
\cite{AhSh}
\begin{equation}
    \label{U.1}
    U^{(1)} (x, y) = - U_0 \ln \left( 1 + \frac{\beta R^2}{x^2 + y^2 + \alpha R^2} \right) \ \ ,
\end{equation}
where $U_0 = 66.6$~eV, $\alpha = 0.48$, $\beta = 1.5$, $R =
0.194$~\AA~(Thomas--Fermi radius). These strings form in the plane
$(100)$ the square lattice with the period $a = a_z / 2\sqrt{2}
\approx 1.92$ \AA , where $a_z = 5.431$ \AA \ is the period of Si
crystal lattice.  Account of the contributions from the eight
closest neighbors to the potential of the given string leads to
the following potential energy of the channeling electron:
\begin{equation}
    \label{U.el}
    U^{(-)} (x, y) = \sum_{i=-1}^1 \sum_{j=-1}^1 U^{(1)} (x-ia, y-ja) + U^{(-)}_0 ,
\end{equation}
where the constant $U^{(-)}_0 = 9.8083$ eV is chosen to achieve
zero potential in the corners of the elementary cell
(figure~\ref{fig.potentials}, left panel).

The positrons can perform axial channeling near [100] direction
due to small potential well formed near the center of the square
cell with repulsive potentials $-U^{(1)}$ in the corners of the
square:
\begin{equation}
    \label{U.pos}
    U^{(+)} (x, y) = -U^{(1)} \left(x-\frac{a}{2}, y-\frac{a}{2} \right)
    -U^{(1)} \left(x-\frac{a}{2}, y+\frac{a}{2} \right) -
\end{equation}
\begin{equation*}
    -U^{(1)} \left(x+\frac{a}{2}, y-\frac{a}{2} \right)
    -U^{(1)} \left(x+\frac{a}{2}, y+\frac{a}{2} \right) - U^{(+)}_0 ,
\end{equation*}
where the constant $U^{(+)}_0 = 7.9589$ eV is chosen to achieve
zero potential in the center of the square elementary cell
(figure~\ref{fig.potentials}, right panel).

\begin{figure}[htbp]
\centering
\includegraphics[width=\textwidth]{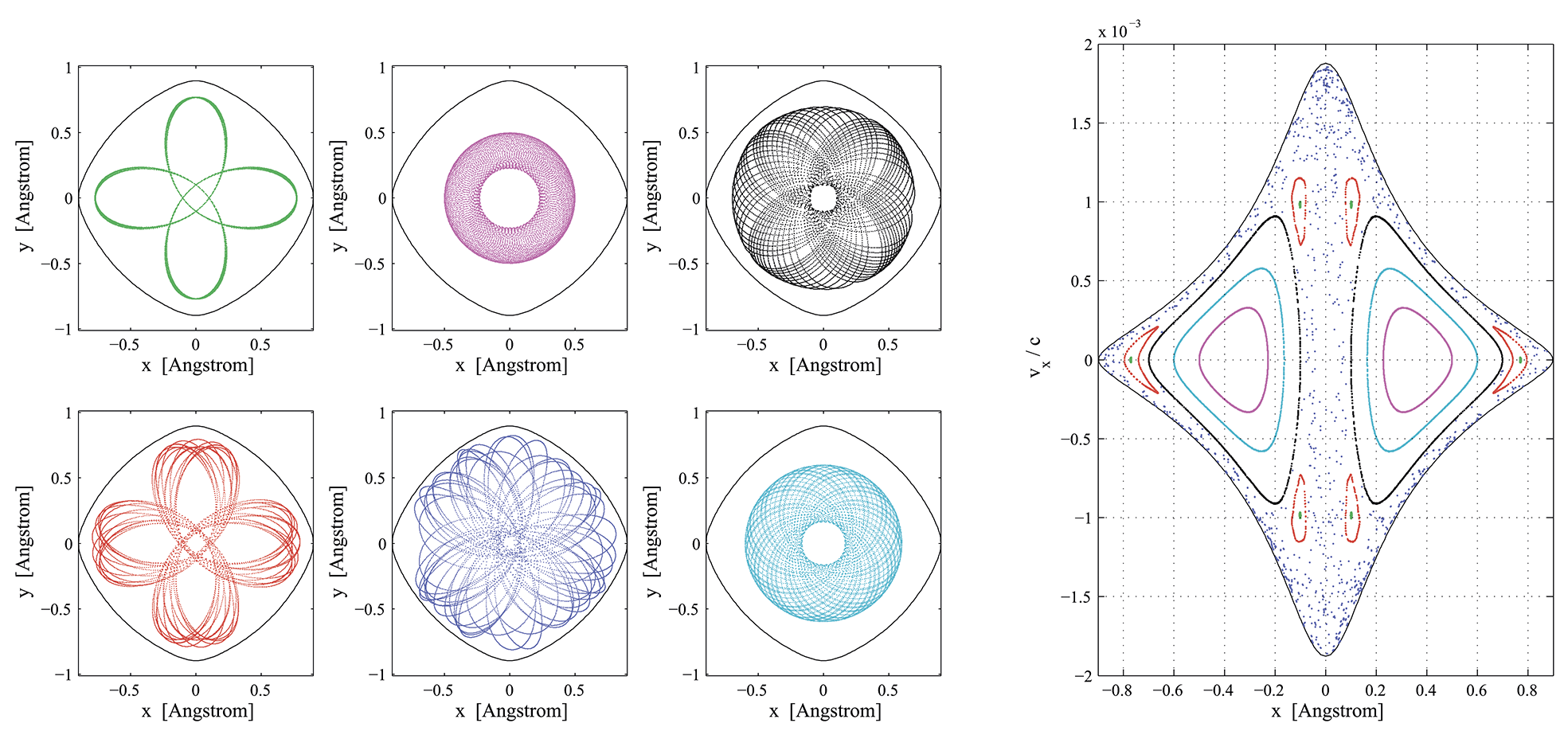} \\
\includegraphics[width=\textwidth]{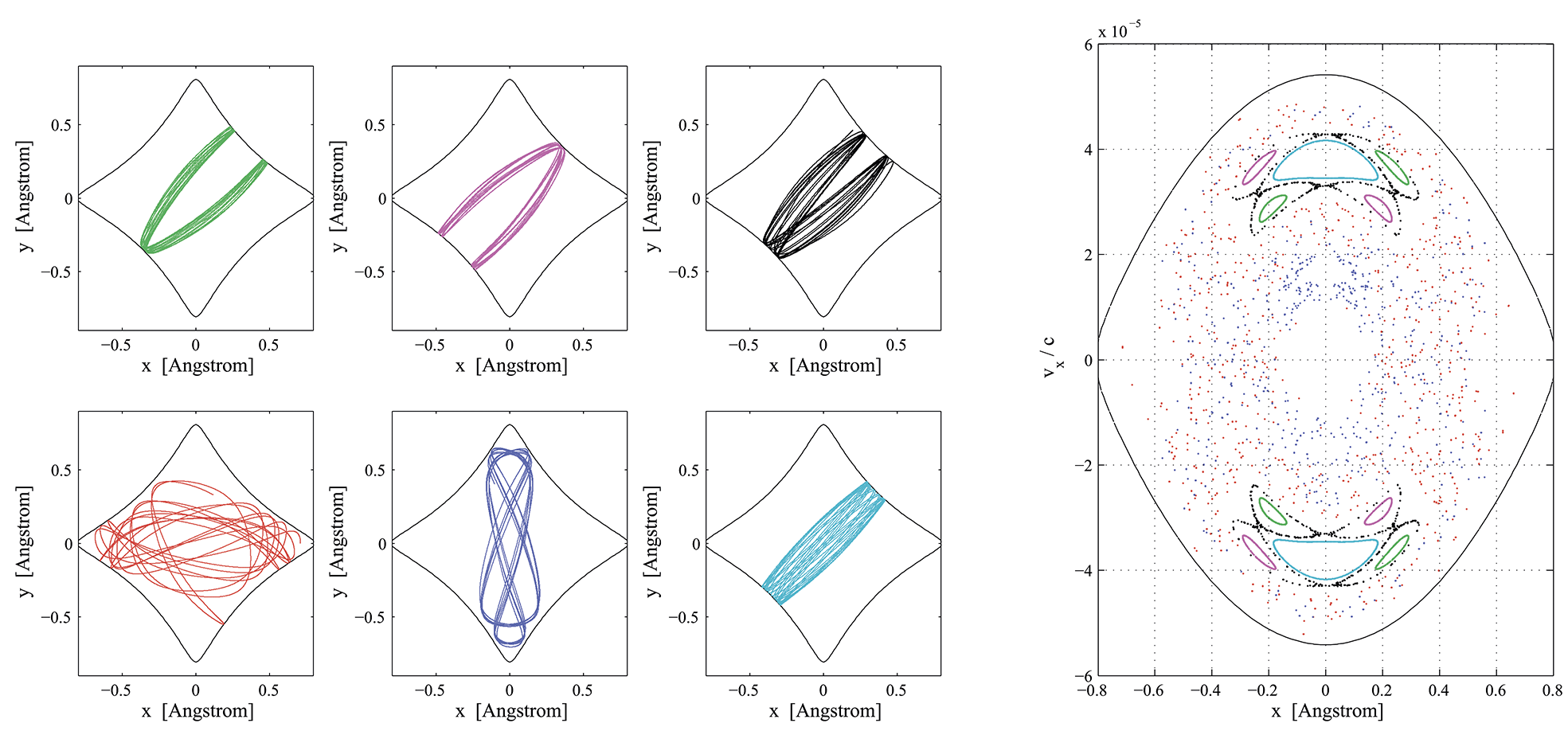}
\caption{\label{Poincare.maps} Typical orbits and Poincar\'e
section of $E_\parallel = 50$~MeV electron channeling in the
potential (\ref{U.el}) with $E_\perp = -2.3957$ eV (upper panel)
and $E_\parallel = 1$~GeV positron channeling in the potential
(\ref{U.pos}) with $E_\perp = 1.4665$ eV (lower panel).}
\end{figure}

So, the electrons channeling along [100] direction move in a
weakly disturbed, almost axially symmetric potential. Their motion
in this potential is regular for the most part of initial
conditions, that is illustrated by Poincar\'e section in figure \ref{Poincare.maps}
(upper panel). In contrast, the dynamics of channeling positrons
is chaotic for the most part of the initial conditions. That is illustrated 
in figure \ref{Poincare.maps} (lower panel).

The electromagnetic transitions from upper to lower energy
levels of the transverse motion (subsection \ref{transitions}) produce the channeling radiation
(CR). We shall calculate CR spectrum using dipole approximation.
The contribution of the given transition $\left| i \right> \to
\left| f \right>$ to the radiation spectrum is described by the
formula (see, e.g., \cite{Baz-Zhev}; the simpler one-dimensional
case of planar channeling is described also in \cite{AhSh}):
\begin{equation}\label{dipole.contribution}
    \frac{d\mathcal E_{fi}}{d\omega} = \frac{\hbar\omega \, dw_{fi}}{d\omega} =
T \, \frac{e^2\omega}{c^3} \, \Omega_{fi}^2 \, \left| \boldsymbol\rho_{fi} \right|^2 \left[ 1 - 2\frac{\omega}{2\gamma^2 \Omega_{fi}}
\left( 1 - \frac{\omega}{2\gamma^2 \Omega_{fi}} \right) \right] \, \Theta (2\gamma^2 \Omega_{fi} - \omega) ,
\end{equation}
where $T$ is the total time of the particle's motion in the
channel (the radiation energy losses all over this time interval
are presumed small), $\gamma$ is the particle's Lorentz factor,
$\Omega_{fi} = (E_\perp^{(f)} - E_\perp^{(i)})/\hbar$ is the
transition frequency, $\Theta(x)$ is the Heaviside step function,
$\boldsymbol\rho_{fi}$ is the dipole moment of the transition,
\begin{equation}\label{dipole}
    \boldsymbol\rho_{fi} = \int \psi_\perp^{(f)} (x,y)^* \, \boldsymbol\rho \, \psi_\perp^{(i)} (x,y) \,dxdy ,
\end{equation}
$\boldsymbol\rho  = x \mathbf e_x + y \mathbf e_y$ is
two-dimensional radius vector in the $(x,y)$ plane.

We shall take into account only the transitions for which the
value (\ref{dipole}) exceeds some threshold, namely
\begin{equation}\label{threshould}
    \left| \boldsymbol\rho_{fi} \right| \geq 10^{-2} \mbox{ \AA} ;
\end{equation}
remember that the elementary cell size over which the integration
is performed in (\ref{dipole}) amounts $a \approx 1.92$ \AA . The
introduction of the threshold permits to exclude from the
consideration the artifacts connected to numerical errors.

The statistical properties of the sequences of energy levels will be considered in subsection \ref{statistics}. For this goal the original set of energy levels has to be unfolded according to the procedure described in \cite{Reichl, Bohigas} in order to get rid of the smooth variations of the levels density from bottom to top of the potential well.

\section{Results and discussion}
\subsection{Channeling electron wave functions structure}

First of all let us consider the low energy ($E_\parallel =
50$~MeV) electron's motion near direction of the atomic string
[100] of the Si crystal.

The potential (\ref{U.el}) within the elementary cell in the (100)
plane is the potential of the single string (\ref{U.1}) weakly
perturbed by the influence from the closest neighbors. Remember
that the motion in the axially symmetric potential of the single
string is integrable: the polar coordinates $r = \sqrt{x^2 + y^2}$ and $\varphi = \arctan (y/x)$
separates, and the Hamiltonian eigenfunctions split into the
products of radial and angular parts. These eigenstates can be
classified by two quantum numbers, the radial $n_r$ and the
orbital $m$. The states of $m=0$ are non degenerated; the states
of $m\not=0$ are twice degenerate. Remember also (see, e.g.,
\cite{Stockmann}) that the eigenfunctions of the real Hamiltonian
without magnetic field and spin (like our (\ref{Hamiltonian}))
always can be chosen real. Hence we choose the functions
\begin{equation}
    \label{psi.cos}
    \rho_{n_r,\, m} (r) \cos (m \varphi)
\end{equation}
and
\begin{equation}
    \label{psi.sin}
    \rho_{n_r,\, m} (r) \sin (m \varphi)
\end{equation}
as the basis functions for $m\not=0$. This choice is highly
demonstrative since the value $m$ in this case manifests itself in
the number of straight nodal lines $\psi(x,y)=0$ travelling through
the origin of coordinates (while the value $n_r$ determines the
number of circular nodal lines with the center in the origin). These lines can be easily seen in black-and-white plots of the eigenfunctions (figure \ref{fig.50.MeV.bw}). Note that the crossing of the nodal lines and the resulting checkerboard-like structure \cite{8} is directly related to separability of the variables in the equation of motion and hence is the characteristic feature of regular quantum systems; this structure is not observed in the chaotic case. 

\begin{figure}[htbp]
\centering
\includegraphics[width=\textwidth]{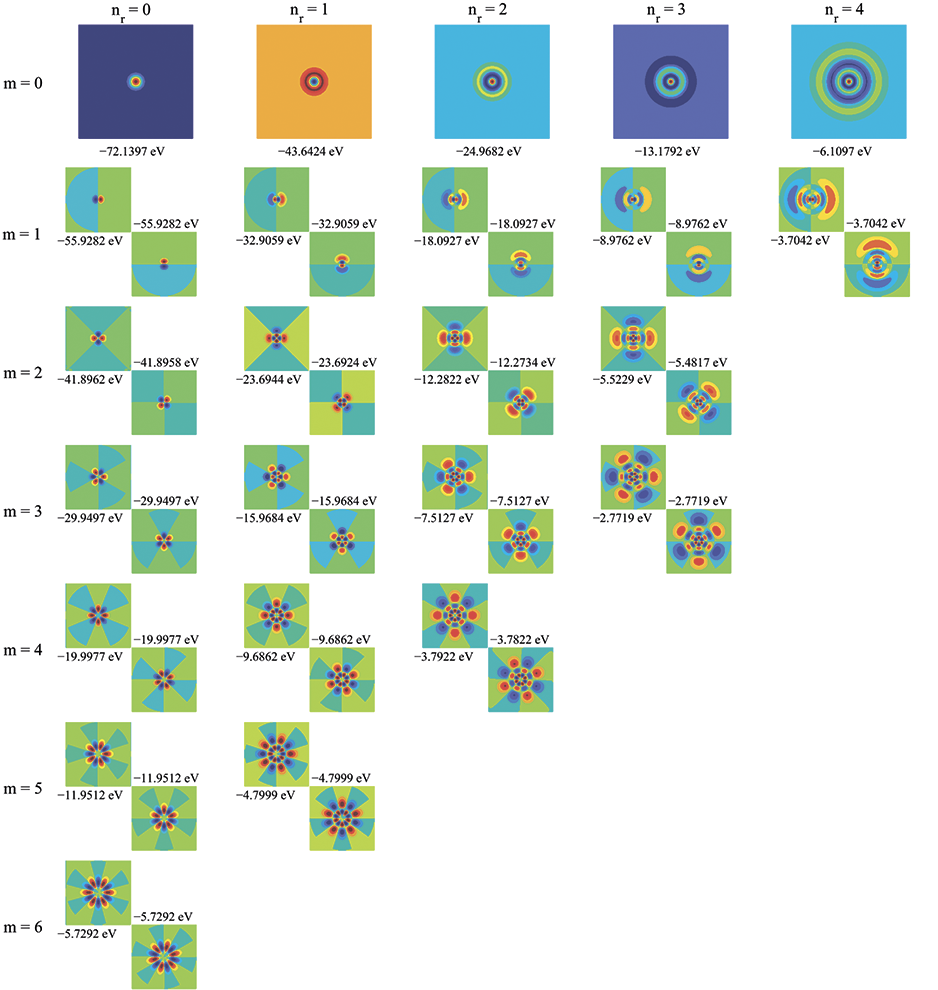}
\caption{\label{fig.50.MeV} The transverse motion eigenfunctions
of $E_\parallel = 50$~MeV electron channeling along [100]
direction of Si crystal.}
\end{figure}

\begin{figure}[htbp]
\centering
\includegraphics[width=\textwidth]{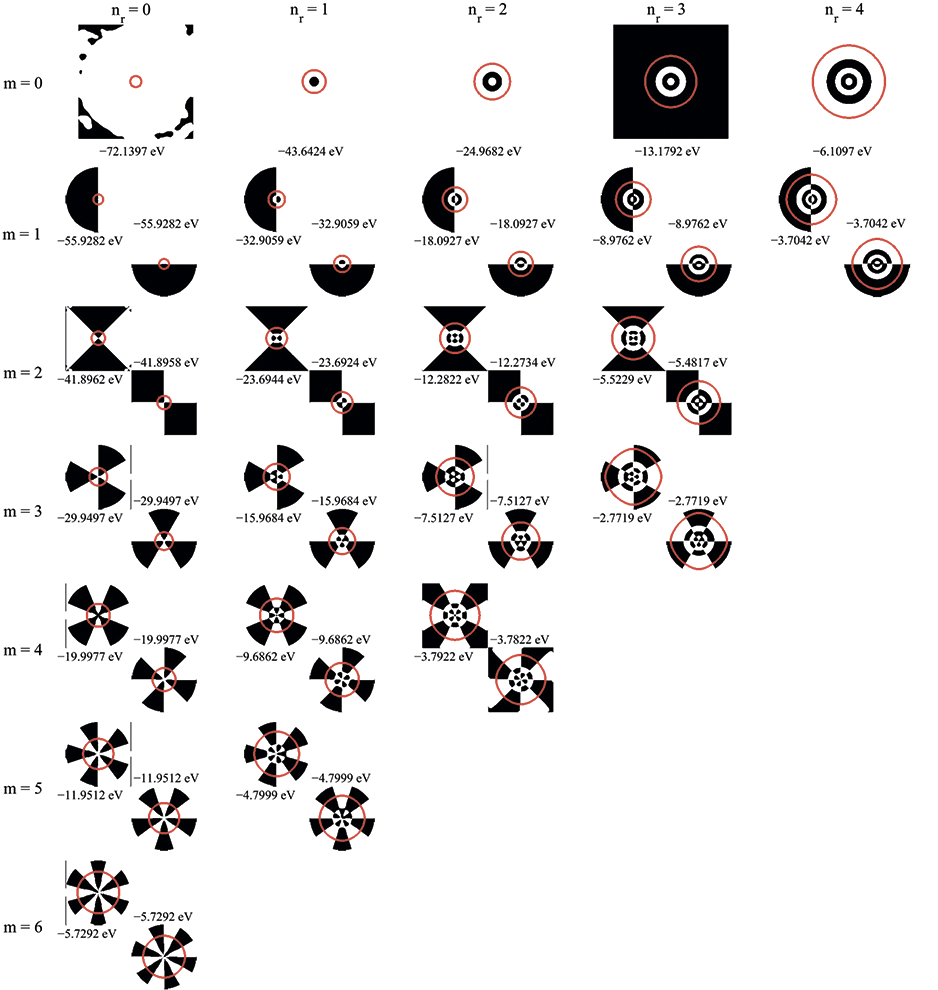}
\caption{\label{fig.50.MeV.bw} The same as in figure \ref{fig.50.MeV} in black and white that allows us to see the zero lines $\psi(x,y) = 0$ of the wave functions easily. Red lines mark the classical border of motion $U^{(-)} (x,y) = E_\perp$.}
\end{figure}

The axial symmetry violation due to the perturbation from the
neighbors partially breaks the degeneracy. The character of this
partial splitting can be predicted using the group theory. The
potential (\ref{U.el}) possesses the symmetry of the square that
is described by the dihedral group $D_4$ (or the group $C_{4v}$
isomorphic to the first one). This group has four one-dimensional
irreducible representations and one two-dimensional one, denoted
$A_1$, $A_2$, $B_1$, $B_2$, $E$, respectively (see, e.g., \cite{LL3, Hamermesh}). It appears 
(see, e.g., \cite{Mathews}) that the states (\ref{psi.cos}) with
$m=4, 8, 12\dots$ are transformed under symmetry transformations of the square according to $A_1$ representation whereas the states (\ref{psi.sin}) with the same
$m$ are transformed according to $A_2$ representation. Hence the perturbation shifts the energy eigenvalues in each of these
state pair by generally speaking the different values that
leads to the energy level splitting. The levels with $m=2, 6,
10\dots$ are split in the same way: the states (\ref{psi.cos}) are
the basic ones for $B_1$ representation whereas the states
(\ref{psi.sin}) form one-dimensional bases for $B_2$
representation.

On the other hand, the pairs of states
(\ref{psi.cos})--(\ref{psi.sin}) with odd $m$ make the bases of
the two-dimensional representation $E$. This means that these wave
functions transform into each other under $D_4$ group
transformations (rotations and reflections), so the perturbation
that possesses $D_4$ symmetry shifts both energy eigenvalues of
the pair by the same amount, e.g. their degeneration conserves.

\begin{figure}[htbp]
\centering
\includegraphics[width=0.49\textwidth]{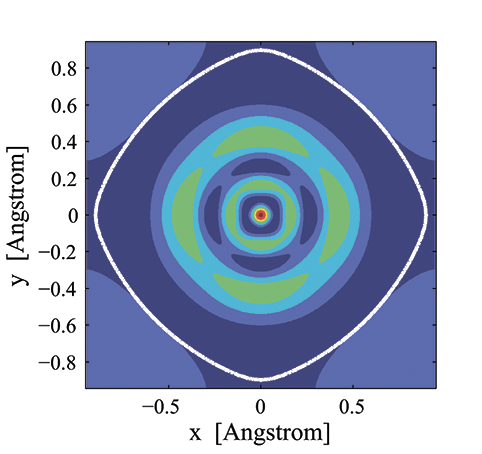} \
\includegraphics[width=0.49\textwidth]{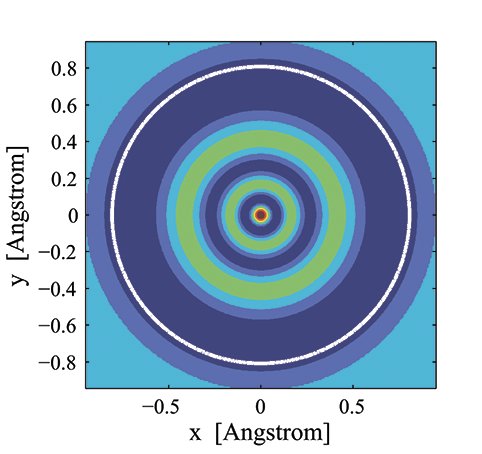}
\caption{\label{fig.50.MeV.highest} The eigenfunction of the
highest energy state $E_\perp = - 2.3957$ eV of the transverse
motion of $E_\parallel = 50$~MeV electron channeling along [100]
direction of Si crystal in the potential (\ref{U.el}) and the
analogous eigenfunction for the motion in the potential of the
isolated string (\ref{U.1}). White lines mark the classical borders of the motion $U(x,y) = E_\perp$.}
\end{figure}

The value of the level splitting could be calculated using the
theory of perturbations. However, it is clear before any
calculations that the splitting would be as high as the influence
of the neighboring strings be strong, and the last one increases
in the upper part of the potential well (\ref{U.el}). This can be
seen in figure \ref{fig.50.MeV}, where the wave functions of all
bound states (except the highest one, pictured in figure \ref{fig.50.MeV.highest}, left panel) of the electron of energy $E_\parallel = 50$~MeV in
the well (\ref{U.el}) are presented with their $E_\perp$ eigenvalues.

Violation of the axial symmetry of the potential manifests itself
in the last case in the wave function structure: instead of the
pure state $\left| n_r = 5, m=0 \right>$ (right panel in figure
\ref{fig.50.MeV.highest}) we see the superposition of the states
with orbital momenta 0 and 4. Note that this feature cannot be
seen in the pattern of the zero lines $\psi (x,y) = 0$.

\subsection{Energy levels and radiation transitions between them}\label{transitions}

\begin{figure}[htbp]
\centering
\includegraphics[width=0.75\textwidth]{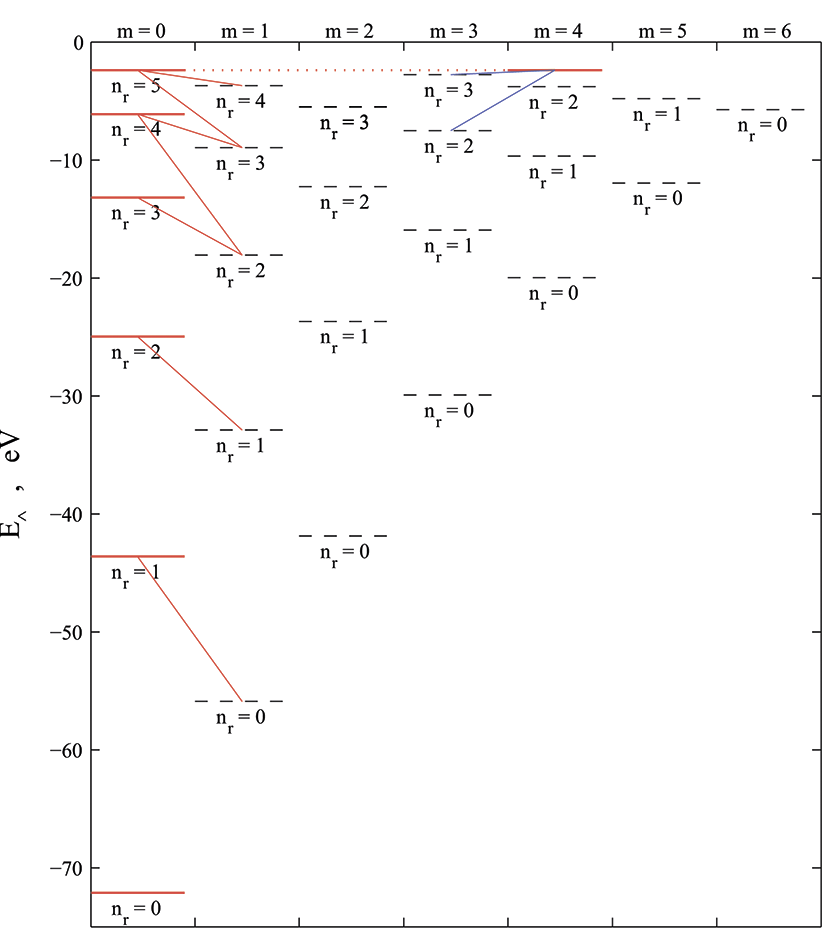}
\caption{\label{fig.transitions.50MeV.9u} Levels of transverse motion energy of the electrons channeling in (\ref{U.el}) potential. Red bars mark the levels populated in the case of zero incidence angle of the beam to [100] axis. The transitions from them that meet the (\ref{threshould}) criterion are shown by red and blue lines, the last ones are the additional transitions possible due to violation of axial symmetry in the potential (\ref{U.el}).}
\end{figure}

The scheme of the transverse motion energy levels for the
channeling electrons is presented in figure
\ref{fig.transitions.50MeV.9u}. The electromagnetic transitions
between them are possible that leads to production of the
channeling radiation (CR).

\begin{figure}[htbp]
\centering
\includegraphics[width=\textwidth]{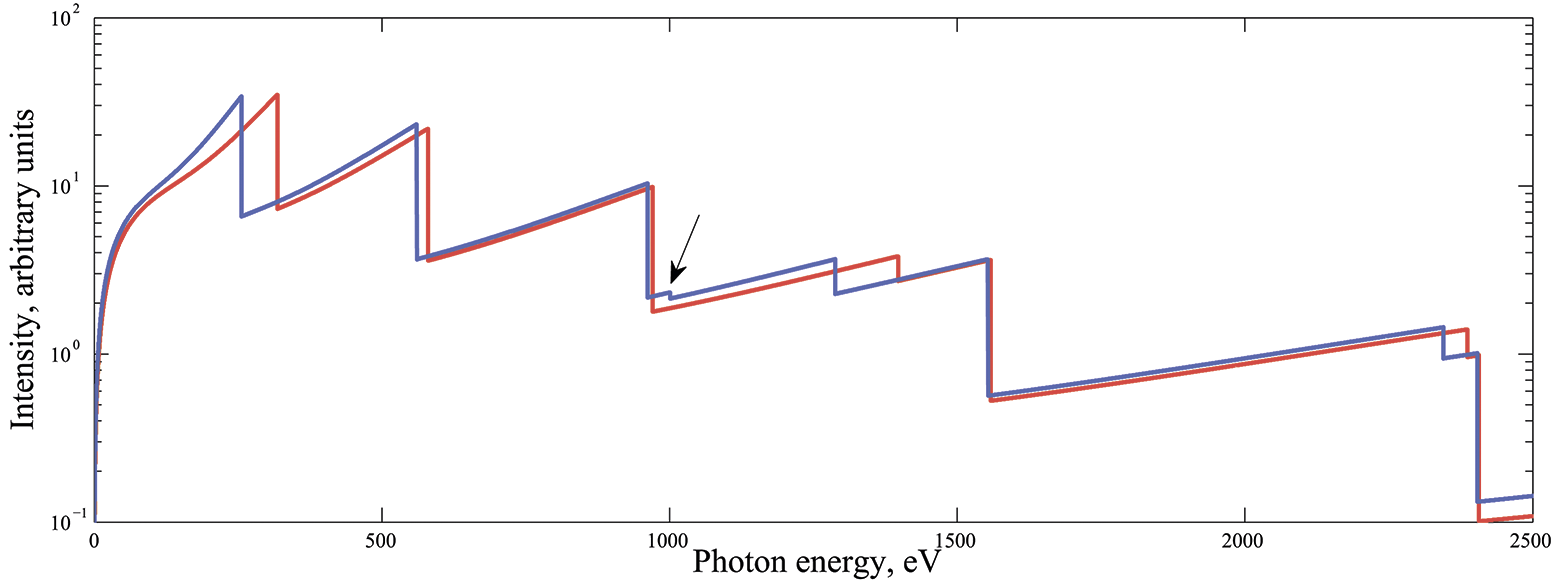}
\caption{\label{fig.CR.spectra.50MeV.9u.vs.1u} Channeling
radiation spectra of $E_\parallel = 50$ MeV electrons in the
potentials (\ref{U.1}) (red line) and (\ref{U.el}) (blue line).}
\end{figure}

The CR spectra in figure \ref{fig.CR.spectra.50MeV.9u.vs.1u} are
computed for the simplest case of thin crystal and zero angle of
the beam incidence to [100] axis. The last condition means that
only the states with $m=0$ are initially populated with electrons
while the first condition allows us to neglect the kinetics of the
levels population during the beam travel through the crystal \cite{Baz-Zhev}.

The character of CR spectrum is determined by the orbital momentum
selection rule: in dipole approximation only the transitions with
$\Delta m = \pm 1$ are permitted. The difference between CR
spectra in the potentials (\ref{U.1}) and (\ref{U.el}) is due to
the fact that outlined above: the highest initially populated
level in the potential of the isolated string (\ref{U.1})  is the
pure state $\left| n_r = 5, m=0 \right>$, so the transitions only
to the states with $m=1$ are permitted. In contrary, the analogous
state in the potential (\ref{U.el}) is the superposition of $m=0$
and $m=4$ states (see figure \ref{fig.50.MeV.highest}), so the
additional transitions from the upper state to the states with
$m=3$ and $m=5$ become possible. These additional transitions are
illustrated in figure \ref{fig.transitions.50MeV.9u} by blue
lines; the probability of one among them, to the state $\left| n_r
= 2, m=3 \right>$, is enough to manifest itself in CR spectrum
(the additional peak due to this transition is pointed in figure
\ref{fig.CR.spectra.50MeV.9u.vs.1u} by the arrow).

So, violation of the axial symmetry of the potential that leads to
chaotization of the motion could manifest itself in additional
peaks in CR spectrum.

\subsection{Energy levels statistics}\label{statistics}

The most pronounced manifestations of chaos in quantum systems are found in the statistical properties of their energy spectra. Consider the distances $s$ between consequent levels in the spectrum of $E_\perp$ eigenvalues. The unfolding procedure \cite{Reichl, Bohigas} leads to dimensionless values of $s$ with the average inter-level spacing $D=1$ for the $E_\perp$ range under consideration. 

The quantum chaos theory predicts (see, e.g., \cite{8, 9, 10, 11, 13}) that the energy levels nearest-neighbor distribution of the chaotic system obeys Wigner function
\begin{equation}
    \label{Wigner}
   p(s) = \frac{\pi}{2} s \exp \left(-\frac{\pi}{4} s^2 \right) 
\end{equation}
while the regular system --- the exponential one
\begin{equation}
    \label{Poisson}
   p(s) = \exp  (-s)
\end{equation}
(frequently referred as Poisson distribution).

\begin{figure}[htbp]
\centering
\includegraphics[width=0.49\textwidth]{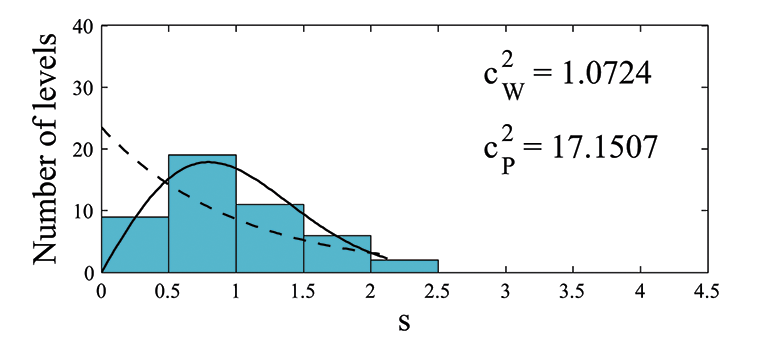} \ 
\includegraphics[width=0.49\textwidth]{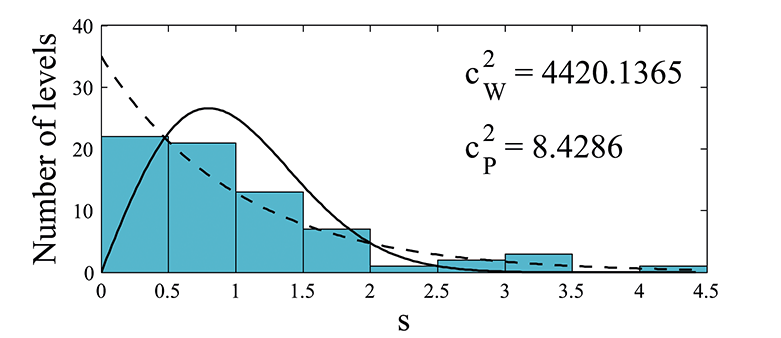} 
\caption{\label{Wigner.vs.Poisson} Nearest-neighbor spacing distribution for the channeling  electrons of $E_\parallel = 5$ GeV in the interval $-3\leq E_\perp \leq -2$ eV (left panel) and $-5\leq E_\perp \leq -3$ eV (right panel).}
\end{figure}

The histograms for the $E_\perp$ level spacing of $E_\parallel = 5$ GeV channeling electron are presented in figure \ref{Wigner.vs.Poisson} for the intervals $-3\leq E_\perp \leq -2$ eV (containing 48 levels) and $-5\leq E_\perp \leq -3$ eV (containing 71 levels). We see that in the first case the distribution is close to Wigner one (that is confirmed by the $\chi^2$ values calculated for both hypotheses, Wigner and Poisson). This result is in agreement with the fact that the domain of chaotic dynamics of the system occupies substantial part of the phase space, $\sim 40\%$ (estimated using Poincar\'e sections).

On the other hand, we see in the second case that the level spacing distribution is closer to (\ref{Poisson}) rather than to (\ref{Wigner}) that is due to mainly regular dynamics of the channeling electron in this $E_\perp$ range (regular trajectories occupy $\sim 90\%$ of the phase space).

\subsection{Channeling positrons wave functions structure}

\begin{figure}[htbp]
\centering
\includegraphics[width=\textwidth]{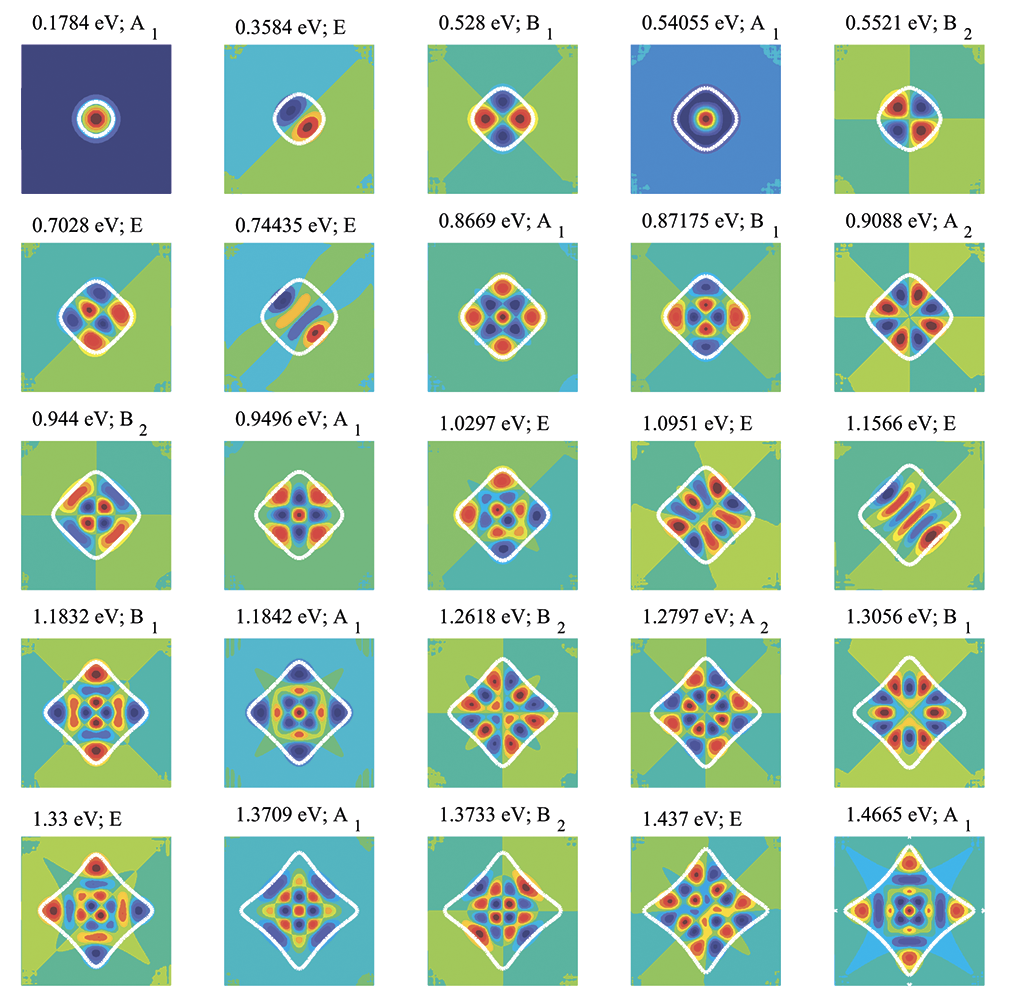}
\caption{\label{fig.1.GeV.pos} The transverse motion eigenfunctions of $E_\parallel = 1$~GeV positron channeling along $[100]$ direction of Si crystal. White lines mark the classical borders of the motion $U(x,y) = E_\perp$.}
\end{figure}

The potential well for the channeling positrons (\ref{U.pos})
could not be considered as a slightly perturbed axially symmetric
well. However, it also possesses the symmetry of the square, hence
the stationary states of the channeling positrons also can be
classified via irreducible representations of $D_4$ group. The
computed eigenfunctions of the channeling positron of the energy
$E_\parallel = 1$~GeV are presented in figure \ref{fig.1.GeV.pos}
(for the twice degenerated states that realize the representation
$E$, only one of the eigenfunctions is pictured; the second one
can be obtained by rotation of the given picture on 90 degrees).

Remember two groups of the qualitative distinctions between the wave functions in the regular and chaotic cases discovered and studied by various authors (see, e.g. \cite{8, 9, 11, NIMB.2016}):
\begin{itemize}
  \item[(i)] the nodal lines of the regular wave function exhibit crossings (in separable case) or very tiny quasi-crossings (in non-separable, but still regular case, see \cite{8}) forming checkerboard-like pattern; the nodal lines of the chaotic wave function form a sophisticated pattern of black and white ``islands'', the nodal lines quasi-crossings have significantly larger avoidance ranges;
  \item[(ii)] near the classical turning line the nodal structure of the regular wave function immediately switches to the straight nodal lines, in the outer domain going to infinity; for the chaotic wave function an intermediate region exists outside the turning line, where some of the nodal lines pinch-off, making transition to the classically forbidden region more graduate and not so manifesting in the nodal structure.
\end{itemize}

\begin{figure}[htbp]
\centering
\includegraphics[width=\textwidth]{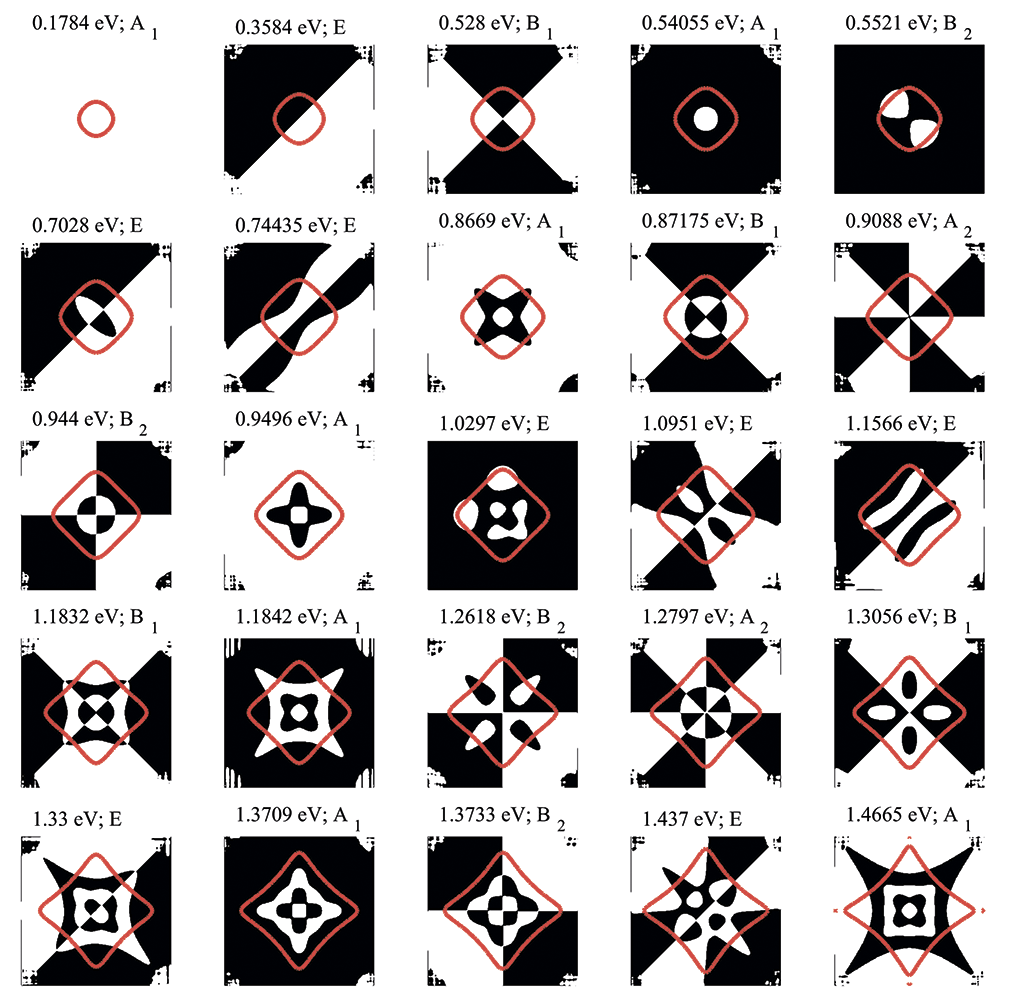}
\caption{\label{fig.1.GeV.pos.bw} The same as in figure \ref{fig.1.GeV.pos} in black-and-white colors.}
\end{figure}

We see both of these features present in black-and-white plots of wave functions of channeling positron (figure \ref{fig.1.GeV.pos.bw}).

\section{Conclusion}

The channeling of electrons and positrons near [100] direction in Si crystal is considered from the quantum-mechanical viewpoint. The energy levels and wave functions of the particle's transverse motion in (100) plane have been computed numerically as well as radiation transitions between formers. We see substantial differences between motion and radiation characteristics in regular and chaotic cases. So, quantum chaos can manifest itself not only in semiclassical case (where the density of energy levels is high), but also in the case of small total number of energy levels.

\acknowledgments

This research is partially supported by the grant of Russian Science Foundation (project 15-12-10019).


\begin{thebibliography}{99}

\bibitem{AhSh} A.I. Akhiezer, N.F. Shul'ga, \emph{High-Energy Electrodynamics in Matter}, Gordon and Breach (1996).

\bibitem{AhSh2} A.I. Akhiezer, N.F. Shul'ga, V.I. Truten', A.A. Grinenko, V.V. Syshchenko, \emph{Physics-Uspekhi} {\bf 38} (1995) 1119.

\bibitem{Ugg} U.I. Uggerh{\o}j, \emph{Rev. Mod. Phys.} {\bf 77} (2005) 1131.

\bibitem{8} R.M. Stratt, N.C. Handy, W.H. Miller, \emph{J. Chem. Phys.} {\bf 71} (1979) 3311.

\bibitem{9} M.C. Gutzwiller, \emph{Chaos in Classical and Quantum Mechanics}, Springer-Verlag (1990).

\bibitem{10}  H.G. Schuster, W. Just, \emph{Deterministic Chaos. An Introduction}, WILEY-VCH Verlag (2005).

\bibitem{11} V.P. Berezovoj, Yu.L. Bolotin, V.A. Cherkaskiy, \emph{Phys. Lett. A} {\bf 323} (2004) 218.

\bibitem{13} M.V. Berry, \emph{Proc. R. Soc. Lond. A} {\bf 413} (1987) 183.

\bibitem{Reichl} L.E. Reichl, \emph{The Transition to Chaos}, Springer (2004).

\bibitem{Poverh.2015} N.F. Shul'ga, V.V. Syshchenko, A.I. Tarnovsky, A.Yu. Isupov, \emph{Journal of Surface Investigation. X-ray, Synchrotron and Neutron Techniques} {\bf 9} (2015) 721.

\bibitem{NIMB.2016} N.F. Shul'ga, V.V. Syshchenko, A.I. Tarnovsky, A.Yu. Isupov, \emph{Nucl. Instrum. Methods in Phys. Res. B} {\bf 370} (2016) 1.

\bibitem{3} M.D. Feit, J.A. Fleck, Jr., F. Steiger, \emph{J. Comput. Phys.} {\bf 47} (1982) 412.

\bibitem{Dabagov3} S. Dabagov, L.I. Ognev, \emph{Nucl. Instrum. and Methods in Phys. Res. B} {\bf 30} (1988) 185.

\bibitem{5} N.F. Shul'ga, V.V. Syshchenko, V.S. Neryabova, \emph{Nucl. Instrum. and Methods in Phys. Res. B} {\bf 309} (2013) 153.

\bibitem{6} N.F. Shul'ga, V.V. Syshchenko, A.Yu. Isupov, \emph{Problems Atom. Sci. Technol.} {\bf 63} 5  (2014) 120.

\bibitem{RREPS15} N.F. Shul'ga, V.V. Syshchenko, A.I. Tarnovsky, A.Yu. Isupov, \emph{J. Phys.: Conf. Series} {\bf 732} (2016) 012028.

\bibitem{Stockmann}  H.-J. St\"ockmann,  \emph{Quantum Chaos. An Introduction}, Cambridge University Press (2000).

\bibitem{LL3} L.D. Landau, E.M. Lifshitz, \emph{Quantum Mechanics. Non-Relativistic Theory}, Pergamon Press (1977).

\bibitem{Hamermesh} M. Hamermesh,  \emph{Group theory and its applications to physical problems}, Addison Wesley Publishing Company (1962).

\bibitem{Mathews} J. Mathews, R.L. Walker, \emph{Mathematical methods of physics}, W.A. Benjamin (1971).

\bibitem{Baz-Zhev} V.A. Bazylev, N.K. Zhevago, \emph{Radiation from fast particles in substance and in external fields}, Nauka (1987), \emph{in Russian}.

\bibitem{Bohigas} O. Bohigas, M.-J. Giannoni, \emph{Lecture Notes in Physics} {\bf 209} (1984) 1.


\end{thebibliography}
\end{document}